\documentclass[conference]{IEEEtran}
\ifCLASSINFOpdf
  \usepackage[pdftex]{graphicx}
  \graphicspath{{/Users/zahmed/mySharedFolder/per/PhD/Latex/IEEEtran/pdf/}{/Users/zahmed/mySharedFolder/per/PhD/Latex/IEEEtran/jpeg/}}
  \DeclareGraphicsExtensions{.pdf,.jpeg,.png}
\else
  \DeclareGraphicsExtensions{.eps}
\fi
\usepackage{algorithm}
\usepackage{algpseudocode}
\usepackage{tabu}
\usepackage{setspace}
\usepackage{makecell}
\usepackage{epsfig}

\usepackage{amsmath}
\usepackage{mathtools}
\usepackage{breqn}
\usepackage{amsfonts}
\usepackage{amssymb}
\begin{document}
%
\title{A Joint Combiner and Bit Allocation Design for Massive MIMO Using Genetic Algorithm}

\author{\IEEEauthorblockN{I. Zakir Ahmed  and Hamid Sadjadpour\\}
\IEEEauthorblockA{Department of Electrical Engineering\\
University of California, Santa Cruz\\
}
\and
\IEEEauthorblockN{Shahram Yousefi}
\IEEEauthorblockA{Department of Electrical and Computer Engineering\\
Queen's University, Canada\\
}}


%


\maketitle

\begin{abstract}
In this paper, we derive a closed-form expression for the combiner of a multiple-input-multiple-output (MIMO) receiver equipped with a minimum-mean-square-error (MMSE) estimator. We propose using variable-bit-resolution analog-to-digital converters (ADC) across radio frequency (RF) paths. The combiner designed is a function of the quantization errors across each RF path. Using very low bit resolution ADCs (1-2bits) is a popular approach with massive MIMO receiver architectures to mitigate large power demands. We show that for certain channel conditions, adopting unequal bit resolution ADCs (e.g., between 1 and 4 bits) on different RF chains, along with the proposed combiner, improves the performance of the MIMO receiver in the Mean Squared Error (MSE) sense. The variable-bit-resolution ADCs is still within the power constraint of using equal bit resolution ADCs on all paths (e.g., 2-bits). We propose a genetic algorithm in conjunction with the derived combiner to arrive at an optimal ADC bit allocation framework with significant reduction in computational complexity.  
\end{abstract}



%
\IEEEpeerreviewmaketitle

\section{Introduction}
In mmWave massive MIMO, a large antenna array is used to form a beam at the receiver. In the Multi-User (MU) case, each user equipment (UE) with its antenna array typically forms a beam to the receiver, which is either a Base Station (BS) or an Access Point (AP). The receiver attempts to spatially decorrelate the signal from $N$ such UEs. A Hybrid precoding/combining is a common architecture used with mmWave massive MIMO transceivers  $\cite{mmPreCom}$, $\cite{SigProc}$, $\cite{hybrid}$. At the receiver, the analog combining will combine the beams and a digital combiner spatially demultiplexes the signals from $N$ users or streams.  The combination of analog and digital precoding and combining will increase the performance of the communication system for a given channel realization $\cite{mmPreCom}$, $\cite{SigProc}$. In another architecture of MU massive MIMO operating at sub-6 Ghz frequencies, the BS with tens to hundreds of antennas would receive signals from $N$ UEs (which typically could be single antenna systems) $\cite{5GBackHaul}$. The BS will demultiplex the signal from $N$ UEs. In both scenarios, the spatial multiplexing of streams increases the capacity of the system linearly with increasing $N$ $\cite{Vishwa}$. 

However, this imposes hardware and particularly power constraints with increasing number of users or independent spatial streams (RF chains) $\cite{SigProc}$. The biggest part of the power consumption on the receiver side is from ADCs. The power consumed by the ADC€™s is exponential in the number of bits (resolution) and directly proportional to the bandwidth of the signal $\cite{SigProc}$, $\cite{Rangan}$ and $\cite{Uplink}$. Prior works analyze the performance of the receivers with 1-bit ADCs to constraint power $\cite{mmPreCom}$, $\cite{5GBackHaul}$, $\cite{SigProc}$. Mixed ADCs with variable bits for each RF chains were proposed in $\cite{Rangan}$, $\cite{Uplink}$, $\cite{VarBitAlloc}$. The intent has been to improve the performance of the receiver by increasing the ADC bit resolution by more than 1 bit at a small sacrifice of power. However the analysis is done with equal bit ADCs on all RF Paths. Distributing the bits unequally across RF paths with power constraint was explored in $\cite{VarBitAlloc}$. In $\cite{VarBitAlloc}$, a bit allocation strategy was arrived in the closed form with a fixed combiner. In our work, we derive the expression for the combiner based on the MMSE criteria which is dependent on the channel. The combiner designed is a function of ADC bit allocation. An optimal bit allocation is obtained using a genetic algorithm together with the help of the combiner designed in the first step. The proposed combiner and bit allocation will be part of digital combining block of the hybrid combining technique for mmWave massive MIMO architecture or simply a linear digital combiner in a massive MU-MIMO framework for sub-6 Ghz frequencies.

We present the simulation results with number of RF paths equal to 8 and 12 using channel models in $\cite{Vishwa}$, $\cite{rapaport}$. We observe that the combined optimal bit allocation and combiner under a given power constraint does not always result in equal distribution of bits to all streams (RF chains); however would depend on the channel $\bold{H}$.

\hfill
 
\section{Signal Model}
In this work, we consider a 'signal' model that captures a variety of MIMO communication systems. Two such examples are: (i) multi-user, multiple-input-multiple-output (MU-MIMO) uplink scenario in which Base Station (BS) with $M$ antennas receive signal from $N$ single-antenna User Equipment (UEs) $\cite{5GBackHaul}$,  $\cite{Uplink}$ and (ii) a mmWave MU-MIMO communication link with hybrid combining where the receiver consists of large number of antennas. Typically, a Uniform Linear Array (ULA) offers combining via analog beamformers and digital combiners $\cite{mmPreCom}$, $\cite{SigProc}$, $\cite{hybrid}$.

In (i), the channel $\bold{H} = \big[ h_{ij} \big]$ is a \begin{math}M\times N\end{math} rich scattering matrix with $h_{ij} \sim \mathcal{N}(0,1)$. The received signal is given by
\begin{equation}\label{eq1}
{\bold{r}} = {\sqrt{p_u}}\bold{H}{\bold{x}}+{\bold{n}},
\end{equation}
where $p_u$ is the average power transmitted per symbol, $M>>N$, $\bold{r}$ is an $M\times1$ receive symbol vector; $\bold{x}$ is an $N\times1$ transmitted signal vector, and $\bold{n}$ is an $M\times1$ noise vector with entries  independent and identically distributed (i.i.d) random variables having complex Gaussian distribution with $\bold{n} \sim \mathcal{CN}(\bold{0},\sigma_n^2\bold{I}_M)$:

In (ii), a mmWave which is either Single-User MIMO (SU-MIMO) or MU-MIMO communication link is considered with hybrid combining. In SU-MIMO case, the communication link is between a single UE and either an AP or BS. In this case, there are $N$ parallel streams of data transmitted and received on $N$ RF paths. In case of MU-MIMO, $N$ UEs talk to an AP/BS receiver. The hybrid combining is divided between the analog and digital domains. The signal model for this system  can be represented as 
\begin{equation}\label{eq2}
{\bold{r}} = {\sqrt{p_u}}{\bold{W}_{\text{rf}}^H}\bold{G}{\bold{x}}+{\bold{W}_{\text{rf}}^H}{\bold{n}},
\end{equation}
where $\bold{r}$ is an $N\times1$ receive symbol vector after analog beam combining from $N$ UEs or parallel data streams, $\bold{W}_{\text{rf}}$ is an $M\times N$ analog combiner,  $\bold{G}$ is an $M\times N$ channel matrix where $M$ is the number of receive antennas at the receiver (usually BS or AP). Let's define $\bold{H}^\prime \triangleq \bold{W}_{\text{rf}}^H\bold{G}$ and ${\bold{n^\prime}} \triangleq {\bold{W}_{\text{rf}}^H}{\bold{n}}$. Channel $\bold{H}^\prime$ can be treated as a matrix whose entries are complex Gaussian random variables $\mathcal{CN}(0,\sigma_n^2)$ and ${\bold{n^\prime}}$ is an $N\times1$ noise vector with entries as i.i.d random variables having complex Gaussian distribution such that $\mathcal{CN}(\bold{0},\sigma_n^2\bold{I}_N)$. Then, \eqref{eq2} can be written as
\begin{equation}\label{eq3}
{\bold{r}} = {\sqrt{p_u}}\bold{H}^\prime{\bold{x}}+{\bold{n^\prime}}.
\end{equation}

It is easy to see that equations~$\eqref{eq1}$ and $\eqref{eq3}$ are analogous to $\bold{H}$ and $\bold{H}^\prime$ being $M\times N$ and $N\times N$ matrices, respectively. We shall consider equation ($\ref{eq1}$) with $M=N$ for our  analysis. The extension of the analysis to $M>N$ is straightforward.
The received symbol vector $\bold{r}$ is digitized using a variable bit quantizer. The quantizer is modeled as an Additive Quantization Noise Model (AQNM) $\cite{Rangan}$, $\cite{Uplink}$. However, when we extend this model for allocating unequal ADC bits (1-4 bits) across $N$ RF paths, the AQNM model $\text{Q}_{\bold{b}} \big( {\bold{r}} \big)$ can be succinctly written as $\cite{VarBitAlloc}$

\begin{equation}\label{eq4}
{\bold{z}}  = \text{Q}_{\bold{b}} \big( {\bold{r}} \big) = \bold{W}_{\alpha}\big( {\bold{b}} \big){\bold{r}}+{\bold{n_q}},
\end{equation}
where $\bold{n}_q$ is the additive quantization noise vector that is uncorrelated with $\bold{r}$ and has Gaussian distribution \cite{Rangan,Uplink,VarBitAlloc},  
${\bold{W}_{\alpha}\left({\bold{b}}\right)} = \text{diag}\left\{\alpha_1,\alpha_2,\alpha_3,...\alpha_N \right\}$, and ${\alpha_i} = 1 - {\beta_i}$. $\beta_i$ is defined as $\beta_i = \frac{\pi\sqrt{3}}{2}2^{-2b_i}$ for a non-uniform MMSE quantizer.
Here, ${\bold{b}} = [b_1 b_2 b_3 .... b_N]^T$ is a vector whose entries $b_i$ indicate the number of bits $b_i$ (on both I and Q channels) that are allocated to the $i^{th}$ ADC  RF path. Based on our proposed bit allocation framework, the number of bits $b_i$ would vary between 1 to 4 depending upon the signal and the channel characteristics.


In the expanded form ${\bold{W}_{\alpha}\left({\bold{b}}\right)}$ can be written as in ($\ref{eq6}$). This approximation holds for $b_i > 5$ and for $b_i \leq 5$, the values are indicated in  Table \ref{betaVal} $\cite{Uplink}$. 
\begin{equation}\label{eq6}
{\bold{W}_{\alpha}\left({\bold{b}}\right)} = \bold{I}_N-\frac{\pi\sqrt{3}}{2} \\
\begin{bmatrix}
2^{-b_1}&0&0&0\\
0&2^{-b_2}&0&0\\
0&0&\ddots&0\\
0&0&0&2^{-b_N}\\
\end{bmatrix}
\end{equation} 

\begin{table}
\begin{center}
\begin{tabu} to 0.5\textwidth { c c c c c c }
 \hline
 $b_i$  & 1 & 2 & 3 & 4 & 5 \\
 \hline
$\beta_i$ & 0.3634 & 0.1175 & 0.03454 & 0.009497 & 0.002499 \\
\hline
\end{tabu}
\vspace{2mm}
\caption{$\beta_i$ for different ADC Quantization Bits $b_i$} \label{betaVal}
\end{center}
\end{table}

\begin{figure}[h]
\begin{center}
\setlength{\unitlength}{4144sp}%
\begingroup\makeatletter\ifx\SetFigFont\undefined%
\gdef\SetFigFont#1#2#3#4#5{%
  \reset@font\fontsize{#1}{#2pt}%
  \fontfamily{#3}\fontseries{#4}\fontshape{#5}%
  \selectfont}%
\fi\endgroup%
\begin{picture}(3856,1029)(1010,-837)
\put(4006,-151){\makebox(0,0)[lb]{\smash{{\SetFigFont{8}{9.6}{\familydefault}{\mddefault}{\updefault}{\color[rgb]{0,0,0}$\bold{C}^H\big( \bold{b} \big)$}%
}}}}
\thinlines
{\color[rgb]{0,0,0}\put(2656,-293){\framebox(768,473){}}
}%
{\color[rgb]{0,0,0}\put(2656,-293){\framebox(768,473){}}
}%
{\color[rgb]{0,0,0}\put(3841,-293){\framebox(767,473){}}
}%
{\color[rgb]{0,0,0}\put(1022,-33){\vector( 1, 0){336}}
}%
{\color[rgb]{0,0,0}\put(2453,-33){\vector( 1, 0){211}}
}%
{\color[rgb]{0,0,0}\put(3422,-33){\vector( 1, 0){422}}
}%
{\color[rgb]{0,0,0}\put(4601,-33){\vector( 1, 0){253}}
}%
{\color[rgb]{0,0,0}\put(2117,-33){\vector( 1, 0){126}}
}%
{\color[rgb]{0,0,0}\put(2251,-16){\line( 1, 0){190}}
}%
{\color[rgb]{0,0,0}\put(2341, 74){\line( 0,-1){190}}
}%
{\color[rgb]{0,0,0}\put(1358,-296){\framebox(767,473){}}
}%
{\color[rgb]{0,0,0}\put(2341,-825){\vector( 0, 1){719}}
}%
\put(1621,-151){\makebox(0,0)[lb]{\smash{{\SetFigFont{8}{9.6}{\familydefault}{\mddefault}{\updefault}{\color[rgb]{0,0,0}$\bold{H}$}%
}}}}
\put(1036, 29){\makebox(0,0)[lb]{\smash{{\SetFigFont{8}{9.6}{\familydefault}{\mddefault}{\updefault}{\color[rgb]{0,0,0}$\bold{x}$}%
}}}}
\put(2476, 29){\makebox(0,0)[lb]{\smash{{\SetFigFont{8}{9.6}{\familydefault}{\mddefault}{\updefault}{\color[rgb]{0,0,0}$\bold{r}$}%
}}}}
\put(3511, 29){\makebox(0,0)[lb]{\smash{{\SetFigFont{8}{9.6}{\familydefault}{\mddefault}{\updefault}{\color[rgb]{0,0,0}$\bold{z}$}%
}}}}
\put(4681, 29){\makebox(0,0)[lb]{\smash{{\SetFigFont{8}{9.6}{\familydefault}{\mddefault}{\updefault}{\color[rgb]{0,0,0}$\bold{y}$}%
}}}}
\put(1527,  8){\makebox(0,0)[lb]{\smash{{\SetFigFont{7}{8.4}{\rmdefault}{\mddefault}{\updefault}{\color[rgb]{0,0,0}Channel}%
}}}}
\put(2874,  8){\makebox(0,0)[lb]{\smash{{\SetFigFont{7}{8.4}{\familydefault}{\mddefault}{\updefault}{\color[rgb]{0,0,0}ADC}%
}}}}
\put(2386,-556){\makebox(0,0)[lb]{\smash{{\SetFigFont{8}{9.6}{\familydefault}{\mddefault}{\updefault}{\color[rgb]{0,0,0}$\bold{n}$}%
}}}}
\put(3970, 29){\makebox(0,0)[lb]{\smash{{\SetFigFont{7}{8.4}{\familydefault}{\mddefault}{\updefault}{\color[rgb]{0,0,0}Combiner}%
}}}}
\put(2870,-151){\makebox(0,0)[lb]{\smash{{\SetFigFont{8}{9.6}{\familydefault}{\mddefault}{\updefault}{\color[rgb]{0,0,0}$\text{Q}_{\bold{b}} \big( {\text{.}} \big)$}%
}}}}
{\color[rgb]{0,0,0}\put(2341,-16){\circle{200}}
}%
\end{picture}%
\caption{Signal Model}
\label{fig:Fig1new.pdf}
\end{center}
\end{figure}
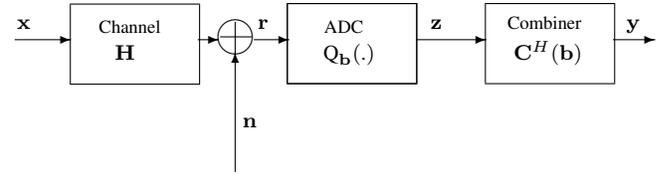

  The quantized signal vector $\bold{z}$ is then combined using a linear combiner $\bold{C}\big(\bold{b}\big)$ as shown in Fig \ref{fig:Fig1new.pdf}.


\subsection{Combiner Design}
We design a combiner $\bold{C} \big( \bold{b} \big)$ such that the mean square error between the transmitted signal vector $\bold{x}$ and the combined output signal vector $\bold{y}$ is minimized.
\begin{equation}\label{eq7}
{\bold{C} \big( \bold{b} \big)_{\text{MMSE}}} = \underbrace{\text{argmin}}_{\bold{C}\big(\bold{b}\big) \in \mathbb{C}^{N \times N}}{\text{E}\lbrace\Vert{{\bold{C}^H\big( \bold{b}\big)}{\bold{z}}-{\bold{x}}}\Vert^2\rbrace}
\end{equation} 

The solution to ($\ref{eq7}$) can be written in the compact form \cite{Vishwa,PhDThesis} as 
\begin{equation}\label{eq8}
{\bold{C} \big( \bold{b} \big)_{\text{MMSE}}}  = {\bold{R}_{{\bold{z}}{\bold{z}}}^{-1}}\big( \bold{b} \big){\bold{R}_{{\bold{z}}{\bold{x}}}}\big( \bold{b} \big),
\end{equation}
where ${\bold{R}_{{\bold{z}}{\bold{z}}}\big( \bold{b} \big)}$ is the covariance matrix of the received and quantized signal vector ${\bold{z}}$, and ${\bold{R}_{{\bold{z}}{\bold{x}}}\big( \bold{b} \big)}$ is the cross-covariance of the transmitted signal vector ${\bold{x}}$ and ${\bold{z}}$. 
The covariance of quantization noise vector ${\bold{n}_q}$ is given by 
\begin{equation}\label{eq9}
{\bold{R}_{{\bold{n}_q}{\bold{n}_q}}} = {\bold{W}_{\alpha}}\big( \bold{b} \big){\bold{W}_{1-\alpha}}\big( \bold{b} \big){\text{diag}\left({p_u\bold{HH}^H+\bold{I}_N}\right)}
\end{equation}

By substituting ($\ref{eq3}$) into ($\ref{eq4}$) and by using the expression for covariance of quantization noise vector, 
  we can compute ${\bold{R}_{{\bold{z}}{\bold{x}}}\big( \bold{b} \big)}$, ${{\bold{R}_{{\bold{z}}{\bold{z}}}}\big( \bold{b} \big)}$, and ${\bold{C} \big( \bold{b} \big)_{\text{MMSE}}}$.
\begin{equation}\label{eq10}
\begin{split}
{\bold{R}_{{\bold{z}}{\bold{x}}}\big( \bold{b} \big)} &= p_u\bold{W}_{\alpha}\big( \bold{b} \big)\bold{H}\\
{{\bold{R}_{{\bold{z}}{\bold{z}}}}\big( \bold{b} \big)} &= p_u\bold{W}_{\alpha}\big( \bold{b} \big)\bold{HH}^H\bold{W}_{\alpha}\big( \bold{b} \big)\\
&+{{\sigma}_n}^2{\bold{I}_N}\bold{W}_{\alpha}\big( \bold{b} \big){\bold{W}_{\alpha}}^T\big( \bold{b} \big) \\
&+{\bold{W}_{\alpha}\big( \bold{b} \big)}{\bold{W}_{1-\alpha}\big( \bold{b} \big)}{\text{diag}\left({p_u\bold{HH}^H+\bold{I}_N}\right)}, 
\end{split}
\end{equation}


\begin{equation}\label{eq12}
\begin{split}
{\bold{C} \big( \bold{b} \big)_{\text{MSE}}} = &\bigg[ p_u\bold{W}_{\alpha}\big( \bold{b} \big)\bold{HH}^H\bold{W}_{\alpha}\big( \bold{b} \big)\\
&+{{\sigma}_n}^2{\bold{I}_N}\bold{W}_{\alpha}\big( \bold{b} \big){\bold{W}_{\alpha}}^T\big( \bold{b} \big) \\
&+{\bold{W}_{\alpha}\big( \bold{b} \big)}{\bold{W}_{1-\alpha}\big( \bold{b} \big)}{\text{diag}\left({p_u\bold{HH}^H+\bold{I}_N}\right)} \bigg]^{-1}\\
&p_u\bold{W}_{\alpha}\big( \bold{b} \big)\bold{H}.
\end{split}
\end{equation}

\subsection{Bit Allocation Formulation}
It is noted that ${\bold{C} \big( \bold{b} \big)_{\text{MSE}}}$ is a non-linear function of $\bold{W}_{\alpha}\big( \bold{b} \big)$. 
We intend to find an optimal $\bold{b}^{*}$ such that ${\bold{C} \big( \bold{b}^{*} \big)_{\text{MMSE}}}$ minimizes the MSE as given in ($\ref{eq7}$), under the power budget constraint $P_{\text{ADC}}$ for ADC. We formulate the cost function $\text{J}\left(\bold{b}\right)$ as 

\begin{equation}\label{eq13}
{\text{E}\lbrace\Vert{e^2}\Vert\rbrace} = \text{J}\left({\bold{b}}\right) = {\text{E}\lbrace\Vert{{\bold{R}_{{\bold{z}}{\bold{x}}}^{\bold{H}}\big( \bold{b} \big)}{\bold{R}_{{\bold{z}}{\bold{z}}}^{-\bold{H}}\big( \bold{b} \big)}{\bold{z}}-{\bold{x}}}\Vert^2\rbrace}.
\end{equation}
The power consumed $\cite{Uplink}$ by a single $b$-bit ADC is given as 
\begin{equation}\label{eq14}
\text{p}\left(b\right) = cf_s2^b,
\end{equation}
where $c$ is the power consumed per conversion step and $f_s$ is the sampling rate in Hz. Given that we have a power budget $P_{\text{ADC}}$, we optimize the function $\text{J}\left(\bold{b}\right)$ under the constraint that the total power consumed by the ADCs with ${\bold{b}}$ bits is less than or equal to $P_{\text{ADC}}$. Let the total power consumed by ADCs using ${\bold{b}}$ be denoted as 
\begin{equation}\label{eq15}
P_{\text{TOT}} = \sum_{i=1}^{N} c{f_s}2^{b_i}.
\end{equation}
We formulate the following optimization problem.
\begin{equation}\label{eq16}
\begin{aligned}
& {\bold{b}^{*}} = \underbrace{\text{argmin}}_{\bold{b} \in \mathbb{I}^{N \times 1}}{\text{E}\lbrace\Vert{{\bold{C}^H}\big( \bold{b} \big){\bold{z}}-{\bold{x}}}\Vert^2\rbrace}\\
& \text{subject to the constraint  } {P_{\text{TOT}}}\leq{P_{\text{ADC}}}
\end{aligned}
\end{equation}

\subsection{Genetic Algorithm for Bit Allocation}
Since the cost function $\text{J}\left(\bold{b}\right)$ is non-linear, we make no assumption about the cost function and solve the optimization problem in ($\ref{eq16}$) using Genetic Algorithm (GA). Given that we need to find an $N$-tuple integer vector as our solution, GA is an attractive choice. We modify the basic framework of the GA described in $\cite{GA}$ to  our problem formulation.

Genetic Algorithms are a class of metaheuristic algorithms that are commonly used to find solutions to optimizations involving non-linear, non-convex cost functions with multiple local minima/maxima or convoluted search spaces that have no closed form representations. GA uses biological principles like mutation and cross-overs to mimic natural selection. When applied to optimization problems, the algorithm selects a set of chromosomes at random into a population set. The chromosomes are the possible solutions to the optimization problem in question. The selected chromosomes are always the ones that adhere to the constraints (for constrained optimization). The fitness test on the chromosomes (evaluation of the cost function) is done on all the chromosomes in the population set. If at any time, a chromosome is found to be fit (passes a minimum or maximum threshold test), the algorithm halts and declare this particular chromosome as the solution.  Otherwise, the GA gets into an iteration loop of generating more chromosomes in the population set by performing cross-over between two or multiple chromosomes in that same set based on a metric called cross-over probability.

The newly formed chromosomes could undergo a mutation based on a mutation probability metric. The new chromosomes thus added into the population set are again evaluated for fitness and the loop continues either till a fit chromosome is found or a decided number of iterations is exhausted. In the scenario where the maximum number of iterations are exhausted without any chromosome passing the test criterion, the most fit chromosome within the population is declared as the solution even though the fitness threshold is not met. The threshold for fitness and the maximum iteration allowed decide the computational complexity of GA $\cite{GA}$.\\

In our proposed GA, as part of initialization we select a set of vectors ${\bold{b}}$ (possible solutions called chromosomes) that adhere to the power constraint and bit allocations as given by 
\begin{equation}\label{eq17}
\begin{split}
B_{\text{set}} = \big\{ &\bold{b}_j = {\big[ b_{j1}, b_{j2}, \dots, b_{jN}  \big]}^T \text{ for } 0 \leq j < 4^N \mid \\
& 1 \le b_{ji} \le 4 \text{ and } \sum_{i=1}^{N} cf_s2^{b_{ji}} \leq P_{\text{ADC}} \big\}
\end{split}
\end{equation}
The initial number of chosen chromosomes is $K$. We call the selected chromosomes set as $Ch_{\text{set}}$.  We also maintain a complimentary set of possible solution vectors $C_{\text{set}}$, that are not  part of $Ch_{\text{set}}$ such that anytime during the GA we have $B_{\text{set}} = Ch_{\text{set}} \cup C_{\text{set}}$ and $Ch_{\text{set}} \cap C_{\text{set}} = \phi$. The fitness $\text{J}\left({\bold{b}}\right)$ is evaluated for each of the ${\bold{b}}$ chromosomes in  $Ch_{\text{set}}$ using ($\ref{eq13}$). We then test if any of the chromosomes have a fitness better than threshold $T$. If so, we exit the GA and declare that chromosome as the solution ${\bold{b}^{*}}$. If none of the chromosomes in $Ch_{\text{set}}$ meets the threshold criteria, we move to the next step in  GA wherein we grow the $Ch_{\text{set}}$ by performing cross-over.  Based on a random measure out of a Bernoulli trial, we pick a new possible solution that gets added to the population. We pick one chromosome from $C_{\text{set}}$ in this fashion for every two distinct chromosomes in $Ch_{\text{set}}$. This is analogous to performing a cross-over of 2 fit chromosomes to create a new chromosome and mutating the same to update the chromosome population $Ch_{\text{set}}$. The cross-over and mutation probabilities are factored in this Bernoulli-trial experiment. The fitness of the chromosomes are evaluated at each iteration when $Ch_{\text{set}}$ is updated . The process is repeated until either a threshold criterion $T$ is reached for fitness or maximum number of iterations $L$ is exhausted, in which case, we pick the most fit chromosome ${\bold{b}}$ as the solution ${\bold{b}^{*}}$. 

\section{Test Setup and Simulation Results}
\begin{figure}[h]
\begin{center}
\setlength{\unitlength}{4144sp}%
\begingroup\makeatletter\ifx\SetFigFont\undefined%
\gdef\SetFigFont#1#2#3#4#5{%
  \reset@font\fontsize{#1}{#2pt}%
  \fontfamily{#3}\fontseries{#4}\fontshape{#5}%
  \selectfont}%
\fi\endgroup%
\begin{picture}(3778,1535)(1,-684)
\put(1493,431){\makebox(0,0)[lb]{\smash{{\SetFigFont{5}{6.0}{\rmdefault}{\mddefault}{\updefault}{\color[rgb]{0,0,0}$\text{Q}_{\bold{b}}\big(\text{.}\big)$}%
}}}}
\thinlines
{\color[rgb]{0,0,0}\put( 16,-421){\vector( 1, 0){284}}
}%
{\color[rgb]{0,0,0}\put( 16,-534){\vector( 1, 0){284}}
}%
{\color[rgb]{0,0,0}\put(318,-616){\framebox(1255,282){}}
}%
{\color[rgb]{0,0,0}\put(583,-80){\vector( 0,-1){254}}
}%
{\color[rgb]{0,0,0}\put(1116,469){\line( 0,-1){ 63}}
}%
{\color[rgb]{0,0,0}\put(1086,438){\line( 1, 0){ 63}}
}%
{\color[rgb]{0,0,0}\put(318,313){\framebox(473,252){}}
}%
{\color[rgb]{0,0,0}\put(1152,431){\vector( 1, 0){283}}
}%
{\color[rgb]{0,0,0}\put(1919,431){\vector( 1, 0){199}}
}%
{\color[rgb]{0,0,0}\put(1444,313){\framebox(471,252){}}
}%
{\color[rgb]{0,0,0}\put(784,431){\vector( 1, 0){285}}
}%
{\color[rgb]{0,0,0}\put(471,313){\vector( 0,-1){647}}
}%
{\color[rgb]{0,0,0}\put(1119, 34){\vector( 0, 1){369}}
}%
{\color[rgb]{0,0,0}\put(3425,147){\vector( 0,-1){339}}
}%
{\color[rgb]{0,0,0}\put(2617,431){\vector( 1, 0){182}}
}%
{\color[rgb]{0,0,0}\put(3515,431){\vector( 1, 0){194}}
}%
{\color[rgb]{0,0,0}\put(2118,313){\framebox(499,256){}}
}%
{\color[rgb]{0,0,0}\put(136,147){\dashbox{57}(3631,692){}}
}%
{\color[rgb]{0,0,0}\put(896,-208){\framebox(472,251){}}
}%
{\color[rgb]{0,0,0}\put( 16,-80){\vector( 1, 0){880}}
}%
{\color[rgb]{0,0,0}\put( 16,431){\vector( 1, 0){302}}
}%
{\color[rgb]{0,0,0}\put(2799,313){\framebox(716,252){}}
}%
{\color[rgb]{0,0,0}\put(206,431){\line( 0, 1){273}}
\put(206,704){\line( 1, 0){2945}}
\put(3151,704){\vector( 0,-1){139}}
}%
{\color[rgb]{0,0,0}\put(1666,-466){\line( 1, 0){720}}
\put(2386,-466){\vector( 0, 1){779}}
}%
{\color[rgb]{0,0,0}\put(1576,-466){\line( 1, 0){ 90}}
\put(1666,-466){\vector( 0, 1){779}}
}%
\put(2118,431){\makebox(0,0)[lb]{\smash{{\SetFigFont{5}{6.0}{\rmdefault}{\mddefault}{\updefault}{\color[rgb]{0,0,0}$\bold{C}^H\big( \bold{b}_{\text{O}} \big)$}%
}}}}
\put(3469,190){\makebox(0,0)[lb]{\smash{{\SetFigFont{5}{6.0}{\rmdefault}{\mddefault}{\updefault}{\color[rgb]{0,0,0}Test Loop}%
}}}}
\put(960,-51){\makebox(0,0)[lb]{\smash{{\SetFigFont{5}{6.0}{\rmdefault}{\mddefault}{\updefault}{\color[rgb]{0,0,0}Generate}%
}}}}
\put(960,-157){\makebox(0,0)[lb]{\smash{{\SetFigFont{5}{6.0}{\rmdefault}{\mddefault}{\updefault}{\color[rgb]{0,0,0}Noise}%
}}}}
\put(381,-460){\makebox(0,0)[lb]{\smash{{\SetFigFont{5}{6.0}{\rmdefault}{\mddefault}{\updefault}{\color[rgb]{0,0,0}Compute  $\bold{C}$ and GA }%
}}}}
\put(412,-553){\makebox(0,0)[lb]{\smash{{\SetFigFont{5}{6.0}{\rmdefault}{\mddefault}{\updefault}{\color[rgb]{0,0,0}to find $\bold{b}_{\text{O}}$}%
}}}}
\put( 16,-24){\makebox(0,0)[lb]{\smash{{\SetFigFont{5}{6.0}{\rmdefault}{\mddefault}{\updefault}{\color[rgb]{0,0,0}$SNR$(dB)}%
}}}}
\put( 16,-363){\makebox(0,0)[lb]{\smash{{\SetFigFont{5}{6.0}{\rmdefault}{\mddefault}{\updefault}{\color[rgb]{0,0,0}$\bold{B}_{\text{set}}$}%
}}}}
\put( 16,-647){\makebox(0,0)[lb]{\smash{{\SetFigFont{5}{6.0}{\rmdefault}{\mddefault}{\updefault}{\color[rgb]{0,0,0}$N$}%
}}}}
\put(471,431){\makebox(0,0)[lb]{\smash{{\SetFigFont{5}{6.0}{\rmdefault}{\mddefault}{\updefault}{\color[rgb]{0,0,0}$\bold{H}$}%
}}}}
\put( 16,489){\makebox(0,0)[lb]{\smash{{\SetFigFont{5}{6.0}{\rmdefault}{\mddefault}{\updefault}{\color[rgb]{0,0,0}$\bold{x}_i$}%
}}}}
\put(3482,-137){\makebox(0,0)[lb]{\smash{{\SetFigFont{5}{6.0}{\rmdefault}{\mddefault}{\updefault}{\color[rgb]{0,0,0}$e$}%
}}}}
\put(2812,406){\makebox(0,0)[lb]{\smash{{\SetFigFont{5}{6.0}{\rmdefault}{\mddefault}{\updefault}{\color[rgb]{0,0,0}Compute MMSE}%
}}}}
\put(3538,489){\makebox(0,0)[lb]{\smash{{\SetFigFont{5}{6.0}{\rmdefault}{\mddefault}{\updefault}{\color[rgb]{0,0,0}$e_i$}%
}}}}
\put(1152,204){\makebox(0,0)[lb]{\smash{{\SetFigFont{5}{6.0}{\rmdefault}{\mddefault}{\updefault}{\color[rgb]{0,0,0}$\bold{n}_i$}%
}}}}
{\color[rgb]{0,0,0}\thicklines
\put(1116,438){\circle{88}}
}%
\end{picture}%
\begin{flushleft}
\begin{spacing}{0.75}
\begin{tabular}{r@{: }l r@{: }l}
\scriptsize{$N$}& \scriptsize{Number of RF paths (8 and 12) in our Tests}\\
\scriptsize{$\bold{x}_i$}&	\scriptsize{$N \times 1$ Symbol vector for $0 \leq i < 400$ Symbols}\\
\scriptsize{$\bold{H}$}& 	\scriptsize{$N \times N$ Channel Matrix}\\
\scriptsize{$\bold{b}_{\text{O}}$}&	\scriptsize{Optimal Bit allocation output of GA}\\
\scriptsize{$\text{Q}_{\bold{b}}\big(\text{.}\big)$}& 	\scriptsize{AQNM as defined by bit allocation $\bold{b}_{\text{O}}$}\\
\scriptsize{$\bold{C}^H\big( \bold{b}_{\text{O}} \big)$}& \scriptsize{Optimal Combiner with bit allocation $\bold{b}_{\text{O}}$. }\\
\scriptsize{$B_{\text{set}}$}& \scriptsize{ \makecell{Set of all possible solutions which adhere to the power constraint \\ $P_{\text{ADC}} = cNB2^{-2}$ (2 bits on all RF paths)}}\\
\scriptsize{$e_i$}& \scriptsize{${\text{J}\big( \bold{b}_{\text{O}}\big) = {\text{E}\lbrace\Vert{{\bold{C}^H \big( \bold{b_{\text{O}}} \big)}{\bold{z_i}}-{\bold{x_i}}}\Vert^2\rbrace}}$ MMSE for symbol $i$}\\
\scriptsize{$e$}& \scriptsize{MMSE averaged over 400 symbols and 100 iterations}
\end{tabular}
\end{spacing}
\end{flushleft}
\caption{Simulation test setup}
\label{fig:Fig2TestSetup.jpg}
\end{center}
\end{figure}
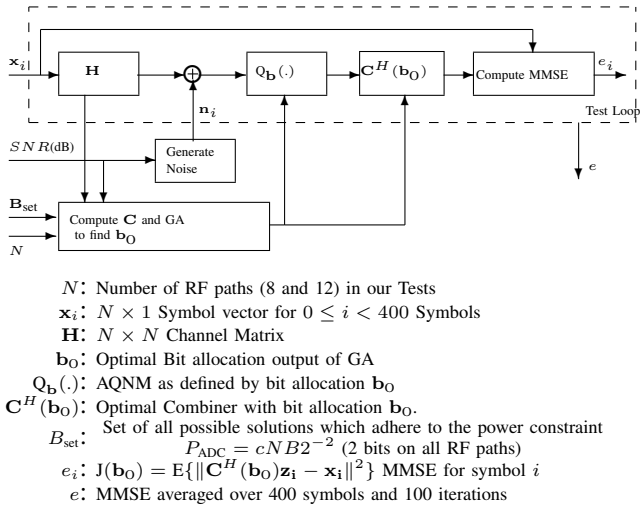

With our Test setup, we simulate $N$= 8 and 12 RF paths at the receiver having ADCs that operate with flexibility to choose bit resolutions between 1 and 4 bits on each RF path. The Channel $\bold{H}$ in this setup is assumed ill-conditioned with condition number greater than 500. We use $N$ = 8 and 12 parallel data streams having 400 symbols each of which is modulated using 64-QAM. The channel is assumed to be stationary over these 400 symbols. The AQNM described in ($\ref{eq4}$) is used to simulate additive ADC quantization noise. The AWGN model is used to simulate different Signal-to-Noise (SNR) conditions. This setup is illustrated in  Fig-$\ref{fig:Fig2TestSetup.jpg}$. The GA parameters selected for our tests are given in Table \ref{tab:GAPar}.

\begin{table}
\begin{center}
\begin{tabu} to 0.5\textwidth { | X[c] | X[c] | X[c] | X[c] | }
 \hline
 Number of RF paths $N$ & Initial Number of Chromosomes $K$ & Maximum Iterations to update chromosomes $L$ & Threshold Test Criteria for Chromosome fitness $T$ \\
 \hline
8  & 64 & 4 & 0.001 \\
\hline
12  & 400 & 4 & 0.001 \\
\hline
\end{tabu}
\vspace{2mm}
\caption{Parameters of GA search} \label{tab:GAPar}
\end{center}
\end{table}

\begin{figure}[!t]
\centering
\includegraphics[width=0.5\textwidth]{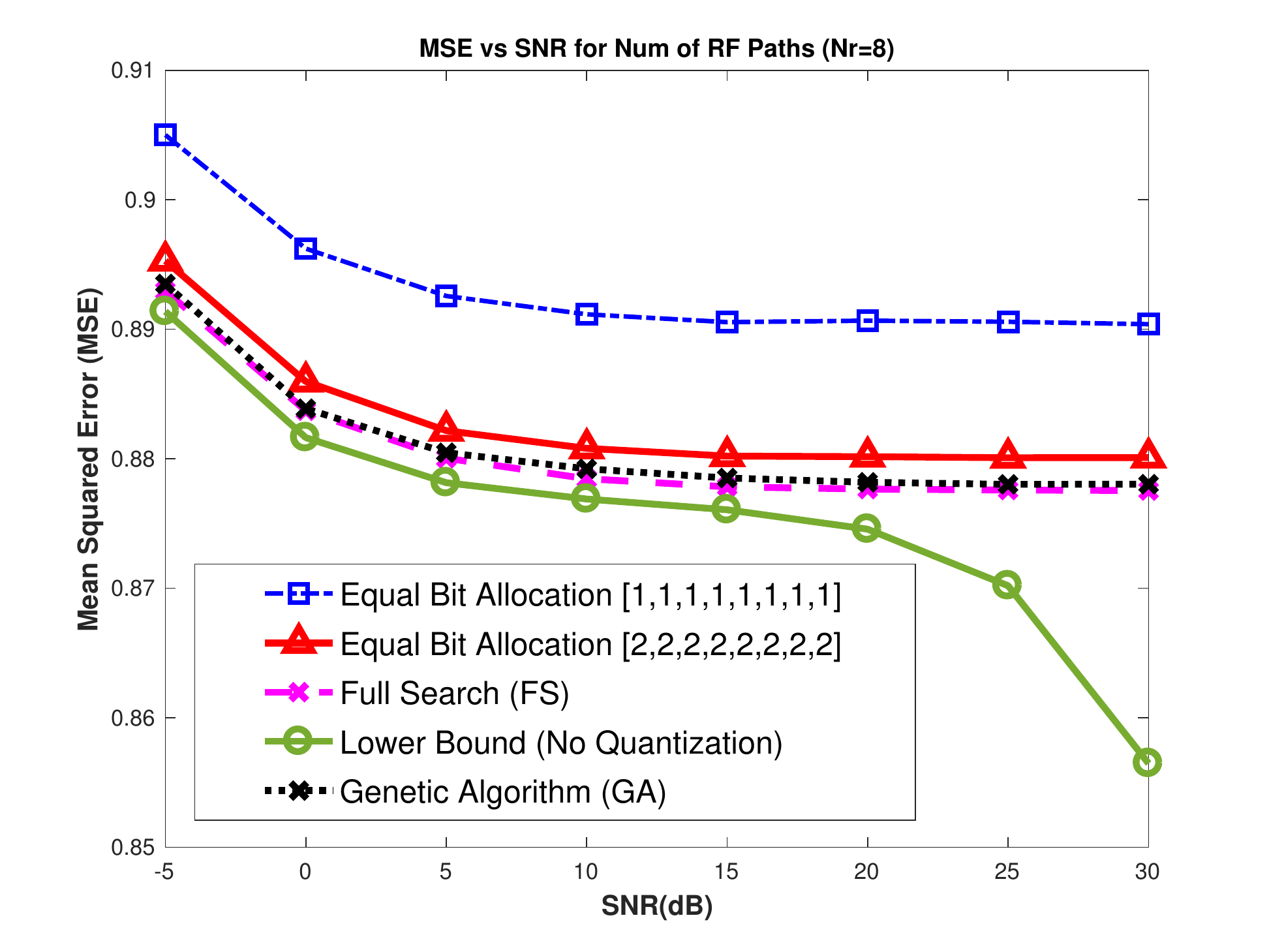}
\caption{MSE vs. SNR for $N=8$}
\label{fig:Nr8_FinalFig.eps}
\qquad
\includegraphics[width=0.5\textwidth]{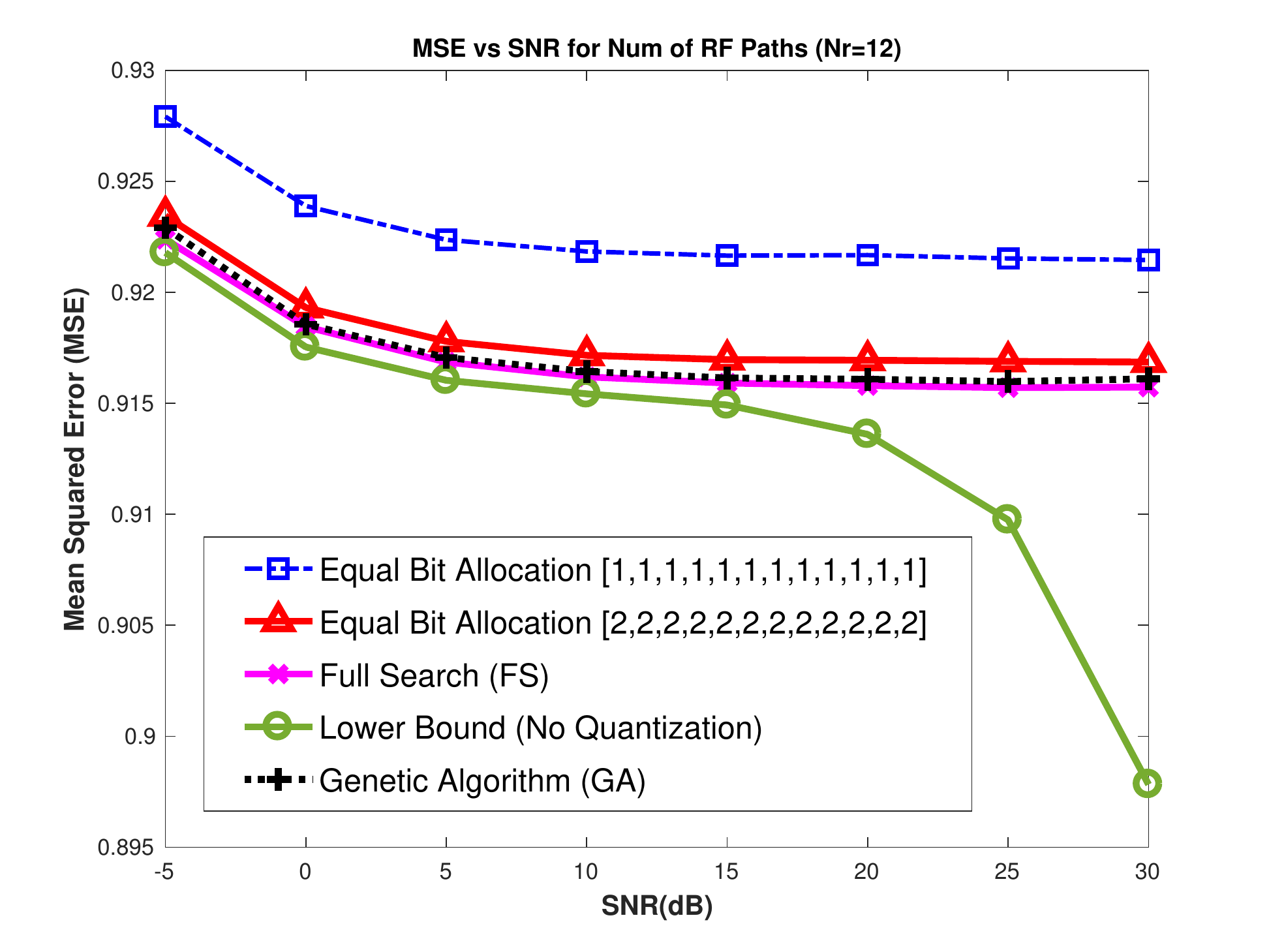}
\caption{MSE vs. SNR for $N=12$}
\label{fig:Nr12_GA_Latest.eps}
\end{figure}

Using this test setup, we run the Full Search (FS) technique and Proposed GA search to find an optimal bit allocation vector for a given channel $\bold{H}$. We set the ADC power budget $P_{\text{ADC}} = cNf_s2^{2}$, that is, the power consumed for having 2-Bit ADCs on all RF paths. Under this constraint, we allow the FS or GA to select ADC bit resolutions between 1 and 4 bits across the RF chains.\\ 
\newline
In FS technique, the cost function $\text{J}\left({\bold{b_j}}\right)$ is computed for every vector $\bold{b}_j$ in the constrained solution set  $B_{\text{set}}$. We then select the $\bold{b}_j$ that yields the minimum $\text{J}\left({\bold{b_j}}\right)$ for a given SNR as the FS solution $\bold{b}_{\text{FS}}\big(\text{SNR} \big)$.  We use this vector to compute the AQNM. Thus, $\bold{b}_{\text{FS}}\big(\text{SNR} \big)$  is computed for different SNRs in the range -5 to 30dB in steps of 5dB. The $\text{MSE}_{\text{FS}}\big(\text{SNR}\big)$ is computed using $\bold{b}_{\text{FS}}\big(\text{SNR}\big)$ for each SNR in the above range using $\eqref{eq13}$. The plot $\text{MSE}_{\text{FS}}\big(\text{SNR}\big)$ vs. SNR thus obtained is shown in pink in the simulation results Fig-$\ref{fig:Nr8_FinalFig.eps}$ and Fig-$\ref{fig:Nr12_GA_Latest.eps}$.
Similarly, with the proposed GA, we find the optimal solution $\bold{b^{*}}\big(\text{SNR}\big)$ at different SNR. Using this solution $\bold{b^{*}}\big(\text{SNR} \big)$,  $\text{MSE}_{\text{GA}}\big(\text{SNR}\big)$ is computed for SNR in the above range. The plot of $\text{MSE}_{\text{GA}}\big(\text{SNR}\big)$ vs. SNR is shown in black in the simulation results of 
Fig-$\ref{fig:Nr8_FinalFig.eps}$ and Fig-$\ref{fig:Nr12_GA_Latest.eps}$.
 
The MSE performance of the MIMO receiver using 2-Bit ADCs on all RF paths is shown as a plot of $\text{MSE}_{\text{2Bits}}\big(\text{SNR}\big)$ vs. SNR in red in Figures  
$\ref{fig:Nr8_FinalFig.eps}$ and $\ref{fig:Nr12_GA_Latest.eps}$. Similarly, the plots in Blue and Green are obtained by having 1-Bit ADCs and infinite bit ADCs (that is, without  quantization errors), respectively.



Table \ref{tab:GATab} describes the computational performance results with the proposed GA and FS Algorithms. Here the computational complexity is measured by the number of evaluations of $\text{J}(\bold{b})$ required to arrive at the optimal solutions ${\bold{b}^{*}}$ and ${\bold{b}_{\text{FS}}}$.

\begin{table}
\begin{center}
\begin{tabu} to 0.5\textwidth { | X[c] | X[c] | X[c] | }
 \hline
 Number of RF paths  & Number of Evaluations of the Cost Function $\text{J}(\bold{b})$ with Full Search & Number of Evaluations of the Cost Function $\text{J}(\bold{b})$ with Genetic Algorithm \\
 \hline
8 & 1878 & 324 \\
\hline
12 & 133253 & 2025 \\
\hline
\end{tabu}
\vspace{2mm}
\caption{Cost function evaluations for full search vs. GA} \label{tab:GATab}
\end{center}
\end{table}

\section{Conclusion}
In this paper, we derive an optimal linear digital combiner $\bold{C}$ for a given channel realization $\bold{H}$ with the MMSE criterion. The MMSE factors in the channel and  AQN for optimal ADC bit allocation. We see that the derived combiner $\bold{C}$ is a function of the variable ADC bit allocation vector ${\bold{b}}$ across the RF paths of the MIMO receiver. We then devise a scheme to search for an optimal bit allocation solution ${\bold{b}^{*}}$  using a computationally efficient GA such that  ${\bold{C}(\bold{b}^{*}})$ minimizes the mean-squared error between the transmitted signal vector $\bold{x}$ and the received, quantized and combined vector $\bold{y}$ under a power constraint. From the simulation results, we see that the optimal bit allocation solution for a given channel $\bold{H}$ and given power budget need not always be uniform and depends on the channel  $\bold{H}$. By using variable-bit allocations for ADCs across RF paths, we have more options to choose the power budget for different channel conditions. This becomes significant as the number of users (streams) increases. Also, we observe from the simulation results that for some channel realizations $\bold{H}$, there exists a different optimal solution $\bold{b}^{*}$ in MSE sense within the solution space, which does not  coincide with the all-one or all-two allocations. This optimal $\bold{b}^{*}$ would not meet the power budget of all-one bit allocation, however we observe substantial MSE improvements over the all-one case at the expense of extra power spending.  The proposed technique can be used for various MIMO architectures. Examples include [i] uplink MU-Massive MIMO systems at sub-6 Ghz frequencies, [ii] mmWave based SU-MIMO with $N$ spatial streams or MU-MIMO with $N$ users within the framework of hybrid precoder/combiner by appropriately taking care of the constraints on the analog precoders and combiners $\cite{SigProc}$. \\

\section*{Acknowledgment}
The authors would like to thank National Instruments for the support extended for this work.  



%

\bibliographystyle{IEEEtran}
\bibliography{myBibTexFile}

\end{document}